\documentclass[review]{elsarticle}
\usepackage{amsmath}
\usepackage{graphicx} 
\usepackage{booktabs}
\usepackage{verbatim}
\usepackage{color} %
\usepackage{bm} %
\usepackage{multirow}%
\usepackage{graphicx, subfig}%
\usepackage{colortbl}%
\usepackage{caption}%
\usepackage{xcolor}%

\usepackage{lineno,hyperref}
\modulolinenumbers[5]

\journal{Journal of \LaTeX\ Templates}

    \makeatletter
    \def\ps@pprintTitle{%
       \let\@oddhead\@empty
       \let\@evenhead\@empty
       \def\@oddfoot{\reset@font\hfil\thepage\hfil}
       \let\@evenfoot\@oddfoot
    }
    \makeatother  

\begin{document}

\begin{frontmatter}

\title{Feature-enhanced Generation and Multi-modality Fusion based Deep Neural Network for Brain Tumor Segmentation with Missing MR Modalities\tnoteref{mytitlenote}}
\tnotetext[mytitlenote]{Accepted by Neurocomputing on September 15, 2021.}



\author[mymainaddress,mysecondaddress,mythirdaddress]{Tongxue Zhou}

\author[mymainaddress,mythirdaddress]{St\'{e}phane Canu}
\author[mysecondaddress,myfourthaddress]{Pierre Vera}

\author[mysecondaddress,mythirdaddress]{Su Ruan\corref{mycorrespondingauthor}}
\cortext[mycorrespondingauthor]{Corresponding author}
\ead{su.ruan@univ-rouen.fr}

\address[mymainaddress]{INSA de Rouen, LITIS -Apprentissage, Rouen 76800, France}

\address[mysecondaddress]{Université de Rouen Normandie, LITIS - QuantIF, Rouen 76183, France}

\address[mythirdaddress]{Normandie Univ, INSA Rouen, UNIROUEN, UNIHAVRE, LITIS, France}

\address[myfourthaddress]{Department of Nuclear Medicine, Henri Becquerel Cancer Center, Rouen, France}

\begin{abstract}
Using multimodal Magnetic Resonance Imaging (MRI) is necessary for accurate brain tumor segmentation. The main problem is that not all types of MRIs are always available in clinical exams. Based on the fact that there is a strong correlation between MR modalities of the same patient, in this work, we propose a novel brain tumor segmentation network in the case of missing one or more modalities. The proposed network consists of three sub-networks: a feature-enhanced generator, a correlation constraint block and a segmentation network. The feature-enhanced generator utilizes the available modalities to generate 3D feature-enhanced image representing the missing modality. The correlation constraint block can exploit the multi-source correlation between the modalities and also constrain the generator to synthesize a feature-enhanced modality which must have a coherent correlation with the available modalities. The segmentation network is a multi-encoder based U-Net to achieve the final brain tumor segmentation. The proposed method is evaluated on BraTS 2018 dataset. Experimental results demonstrate the effectiveness of the proposed method which achieves the average Dice Score of 82.9, 74.9 and 59.1 on whole tumor, tumor core and enhancing tumor, respectively across all the situations, and outperforms the best method by 3.5\%, 17\% and 18.2\%.
\end{abstract}

\end{frontmatter}

\section{Introduction}
\label{sec1}
Brain tumor is a growth of cells in the brain that multiplies in an abnormal, uncontrollable way. Gliomas are one of the most common types of brain tumors with different grades according to tumor malignancy and the rate of growth. A higher grade is usually more aggressive and more likely to grow quickly. According to National Brain Tumor Society (NBTS)\footnote{https://cbtrus.org/}, an estimated 25,130 people will be diagnosed as malignant brain tumors in the United States in 2021 \cite{central2007primary}. Therefore, early diagnosis of brain tumors plays an important role in clinical practice and treatment planning. 

\begin{figure}[!t]
\centering
\centerline{\includegraphics[width=\columnwidth]{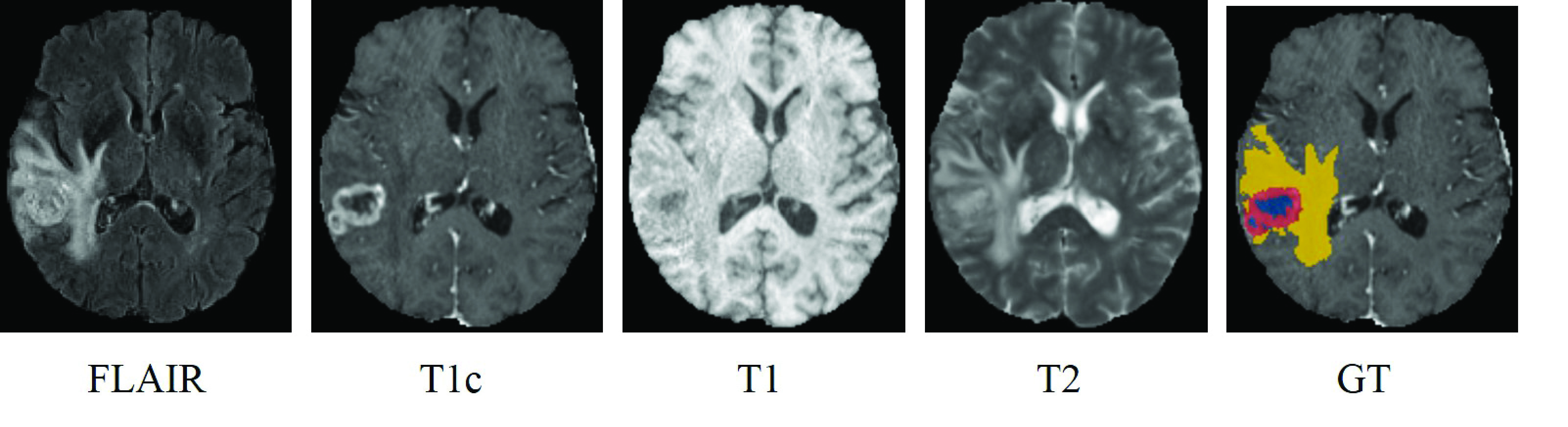}}
\caption{A training sample from BraTS 2018 dataset. From left to right: Fluid Attenuation Inversion Recovery (FLAIR), contrast enhanced T1-weighted (T1c), T1-weighted (T1), T2-weighted (T2) images, and the ground-truth (GT). Net\&ncr is shown in blue, edema in yellow and enhancing tumor in red. Net refers non-enhancing tumor and ncr necrotic tumor.}
\label{fig1}
\end{figure}

Magnetic Resonance Imaging (MRI), as a non-invasive and good soft tissue contrast imaging modality, is commonly used in neurology to visualize the brain. Because it can provide the detailed characteristics about the shape, size, and localization of brain tumors \cite{liang2000principles,bauer2013survey,drevelegas2010imaging, bakas2017advancing, joya2021review, tiwari2020brain}. There are four commonly used sequences: T1-weighted (T1), contrast-enhanced T1-weighted (T1c), Fluid Attenuation Inversion Recovery (FLAIR) and T2-weighted (T2) images. In this work, we refer to different sequences of images as modalities. Different MRI modalities can highlight different sub-regions, which can provide the complimentary information to analyze the brain tumor. To make it clear, we visualize a case from BraTS 2018 dataset in Figure \ref{fig1}. It can be observed that the FLAIR highlights the whole tumor region, and T1c provides more information about tumor core. Intuitively, it's possible to obtain the best segmentation result using full modalities. However, it's common to have one or more missing modalities in clinical practice, due to motion artefacts, scan corruption and limited scan time. 

To this end, we propose a feature-enhanced generation and multi-modality fusion based deep neural network for brain tumor segmentation with missing MR modalities. The preliminary conference versions appeared at 2020 International Conference on Medical Image Computing and Computer-Assisted Intervention (MICCAI) \cite{zhou2020brain} and the International Conference on Pattern Recognition (ICPR) \cite{zhou:hal-03087220}, respectively. They focus on utilizing the multi-source correlation to do the segmentation on missing modalities and full modalities, respectively. The MICCAI version is published in IEEE Transactions on Image Processing (TIP) \cite{zhou2021latent}. In this paper, we only replaced missing modalities by the existing ones. This was a weak point of the method. In this new work, we have developed a new framework to generate the feature map of the missing modalities under correlation constraint. The whole network is completely changed. Moreover, more comparison experiments and parameter settings are addressed. This new work has improved the results of the previous works. The main contributions of our approach are:


(1) A novel feature-enhanced generation and multi-modality fusion based deep neural network is proposed to address the brain tumor segmentation in the case of missing MR modalities. First, an auto-encoder based generator is introduced to generate the feature-enhanced missing modality. Then, a correlation constraint block is proposed to exploit the multi-source correlation between modalities to make the network extracting principally the correlated latent features. Finally, a multi-encoder based U-Net with the correlation constraint of multi-modalities is proposed to do the final segmentation.

(2) To exploit the multi-source correlation between MR modalities, we first designed a dedicated Correlation Parameter Estimation Module (CPEM) to learn the correlated weight parameters for each modality. And then we introduced a Linear Correlation Expression Module (LCEM) to form the correlated representation for each modality. Finally, a novel Correlation Constraint Loss (CCL) is employed to ensure that each modality satisfies the multi-source correlation with other modalities.

(3) The proposed correlation constraint block can guide, on the one hand, the generator to synthesize the feature-enhanced modality via satisfying the multi-source correlation with other available modalities; On the other hand, the following segmentation network to learn useful feature representations so as to achieve a better segmentation performance.

(4) The experimental results evaluated on the public multi-modal brain tumor segmentation dataset demonstrated the effectiveness of each proposed components, and the proposed method can obtain the significant improvements compared with the baseline methods and state-of-the-art methods.

The paper is organized as follows: Section \ref{sec2} reviews the previous work, Section \ref{sec3} offers an overview of this work and details of our network architecture. Section \ref{sec4} describes experimental setup and used datasets. Section \ref{sec5} presents the experimental results. Section \ref{sec6} concludes this work.

\section{Related Works}
\label{sec2}
Brain tumor segmentation in MRI plays an important role in the community of medical science, which has many applications in neurology such as quantitative analysis, operational planning, and functional imaging \cite{wadhwa2019review}. However, it's still a challenging task due to some limitations, such as the complex brain anatomy structure, various shapes, the texture of gliomas, and the low contrast of MR images \cite{zhou2019review, kamnitsas2017efficient}. To address these issues, a multitude of brain tumor segmentation approaches have been proposed in the last decades. It can be generally categorized into two groups, conventional approach and deep learning based approach. There exists some successful conventional approaches, such as Gaussian copula based Bayesian method \cite{lapuyade2017segmenting}, kernel feature selection \cite{zhang2011kernel}, belief function based fusion strategy \cite{lian2018joint}, random forests \cite{zikic2012decision, csaholczi2020brain}, conditional random fields \cite{yu2018semi} and support vector machines \cite{bauer2011fully, krishnakumar2021effective}. Although these methods achieved good performance, they usually have small number of parameters that are insufficient to capture the complex features of brain tumor.

Recently, as a powerful alternative for feature learning, deep learning based approaches have attracted much attention in the field of brain tumor segmentation. For example, Myronenko et al. \cite{myronenko20183d} proposed a 3D multimodal brain tumor segmentation network, which integrated a variational auto-encoder branch to the decoder to help with the limited training data. Chen et al. \cite{chen2019dual} proposed a dual-force training strategy to explicitly encourage deep models to learn high-quality multi-level features for brain tumor segmentation. Chen et al. \cite{chen2020brain} proposed a novel deep convolutional neural network which combines symmetry information to automatically segment brain tumors. Ding et al. \cite{ding2020multi} proposed a multi-path adaptive fusion network to enhance entire feature hierarchy for multi-modal brain tumor segmentation. To handle the class imbalance problem, Zhou et al. \cite{zhou2020one} proposed a light-weight model, OM-Net for brain tumor segmentation, which requires only one-pass computation to perform coarse-to-fine segmentation.

The approaches mentioned above require the complete set of the modalities. However, the full modalities are not always available in clinical practice. It is highly desirable to design an automatic brain tumor segmentation approach to tackle with the missing modality problem. Recently, many related studies have emerged. We generally classify them into three groups: (1) taking all possible combinations of the modalities into account and then training a model based on it, while this method requires a large amount of data and it is time-consuming. (2) synthesizing missing modalities and then use the complete modalities to do the segmentation. This method can not only synthesize the missing modality but also obtain the segmentation results. Generative Adversarial Network (GAN) \cite{goodfellow2014generative, nie2018medical, yu2019ea} is an effective approach for image synthesis. However, the training of a 3D GAN is highly unstable and difficult to converge. In addition, it is difficult to put the generation of a 3D image by GAN and the image segmentation in a same architecture. In the literature \cite{xia2020recovering}, these two tasks are performed separately. In this paper, we include both tasks in the same framework that allows to optimize the generation and segmentation. (3) fusing the available modalities in a latent space to learn a shared feature representation, then project it to the segmentation space \cite{havaei2016hemis, lau2019unified, chartsias2017multimodal, dorent2019hetero}. This approach is more efficient than the first two approaches, since it doesn't need to learn a number of possible subsets of the multi-modalities. 
 
Currently, the approaches based on exploiting latent feature representation for missing modalities become prominent. The early network architecture designed for missing modalities is from HeMIS \cite{havaei2016hemis}. Independent feature maps are first extracted by independent convolutional network for each modality. Then, they are fused via computing the mean and variance for the final segmentation prediction. Similarly, Lau et al. \cite{lau2019unified} proposed a U-Net based network named Unified Representation Network (URN) to combine the independent features by calculating the mean to obtain the unified representation for the final segmentation. To further enhance the modality-invariance of latent representations, Chartsias et al. \cite{chartsias2017multimodal} proposed to minimize the L1 or L2 distance of features from different modalities. Since different MRI modalities have different intensity distributions, using arithmetic operations, such as mean and variance or simply encouraging the features from different modalities to be close under L1 or L2 distance, could not guarantee the network can learn a shared latent representation. To this end, Chen et al. \cite{chen2019robust} introduced the feature disentanglement to tackle the missing data problem. Dorent et al. \cite{dorent2019hetero} proposed a Hetero-Modal Variational Encoder-Decoder (U-HVED) network to use multi-modal variational auto-encoders to embed all available modalities into a shared latent representation, and the experimental results demonstrated that it can outperform HeMIS. Furthermore, Shen et al. \cite{shen2019brain} designed a domain adaptation model to adapt feature maps from missing modalities to the one from full modalities. Hu et al. \cite{hu2020knowledge} employed the generalized knowledge distillation to transfer knowledge from a multi-modal segmentation network to a mono-modal one to achieve the brain tumor segmentation in the case that only one modality is available. Zhu et al \cite{zhu2021brain} proposed a cascade supplement module to first generate shared features for missing modalities and then use squeeze and excitation to fuse the generated features and real features to achieve the segmentation.

\section{Method}
\label{sec3}
\begin{figure*}[!t]
\centering 
\includegraphics[width=0.8\textwidth]{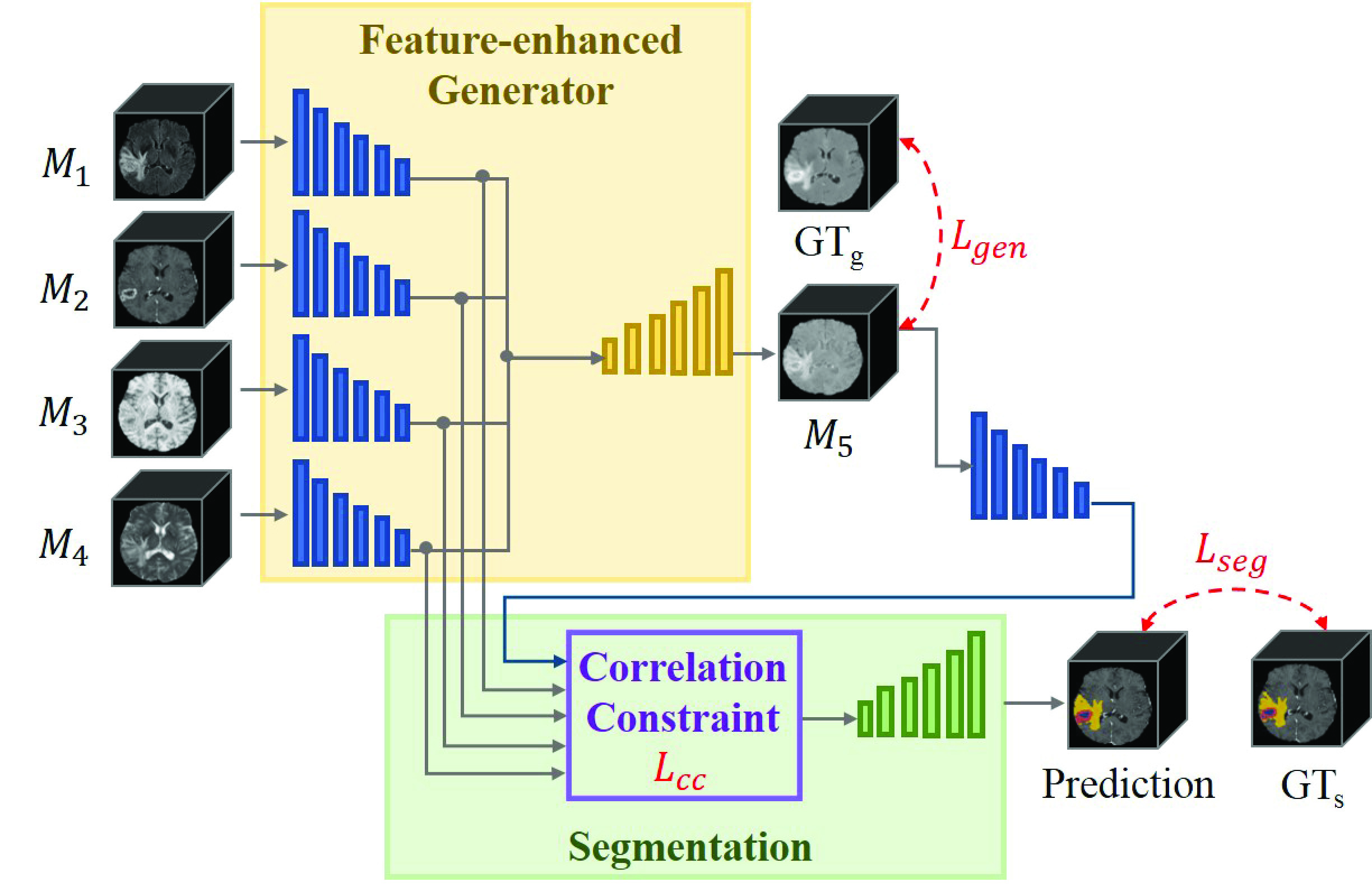}
\caption{An overview of our network, consisting of a Feature-enhanced Generator (FeG), a Correlation Constraint (CC) block and a segmentation network. The feature-enhanced generator utilizes the available modalities to generate a feature-enhanced missing modality $M_5$. Then the complete modalities are input to the segmentation network for the final prediction. In addition, the correlation constraint block are used to guide both the feature-enhanced generator and segmentation network via exploiting the multi-source correlation.}
\label{fig2}
\end{figure*}

The overview of our method is depicted in Figure (other figures also) \ref{fig2}. First, the feature-enhanced generator takes the available modalities as inputs and generates a feature-enhanced modality $M_5$, which is an average of the missing modalities, to form the new complete modalities. At the endpoints of the encoders, the feature representations of each modality are extracted while the feature representation of $M_5$ is extracted by an independent encoder, which has the same architecture as the encoder of the feature-enhanced generator. Following that, a Correlation Constraint (CC) block is used to exploit the multi-source correlation between modalities. On the one hand, it can provide a constraint for the feature-enhanced generator. On the other hand, it can guide the segmentation network to learn the effective feature representations. Specifically, if the synthesized modality is not well generated, then the correlation between the synthesized modality and the available ones will be weak. Thus, through the loss function, the network iteratively generates an increasingly satisfactory feature modality with a high correlation from available modalities. Finally, a segmentation decoder is applied to do the final segmentation.

\subsection{Motivation of the proposed method}
In clinical diagnosis and treatment planning, the patient undergoes multimodal MRI scans because each modality can provide specific information. Our hypothesis is that there is some correlation among multimodal MRIs in the tumor regions. We take an example from BraTS 2018 dataset to present joint intensity distributions of the MR images. From Figure \ref{fig4}, we can observe that there is a strong correlation in intensity distribution between each pair of modalities. To this end, it's reasonable to assume that a strong correlation also exists in the latent representation between modalities \cite{lapuyade2017segmenting}. Therefore, we introduce a Correlation Constraint (CC) block to discover the multi-source correlation between modalities. It can help to recover the features when the modalities are absent. In general, the proposed CC block can be adapted to other computer vision fields. However, the proposed method will depend on the specific image modalities and the relationship among them. Our method is limited to linear correlation. More investigations will be needed for the more complicated correlation situation.

\subsection{Modality dropout} 
\label{sec3.0}
Conventional dropout is a regularization technique for reducing over-fitting, it randomly drops out the units in a neural network. It's intuitive that missing modality will definitely diminish the segmentation performance. Inspired by dropout technique, to make the network robust to the missing modality, we adopt the modality dropout strategy. We randomly zeros out any number of the modality for each sample during the training iterations. To this end, the network can learn to compensate the missing modalities during training, and the performance will not degrade gravely during the test.

\subsection{Synthesizing the missing modality} 
\label{sec3.1}
\begin{figure*}[!t]
\includegraphics[width=\textwidth]{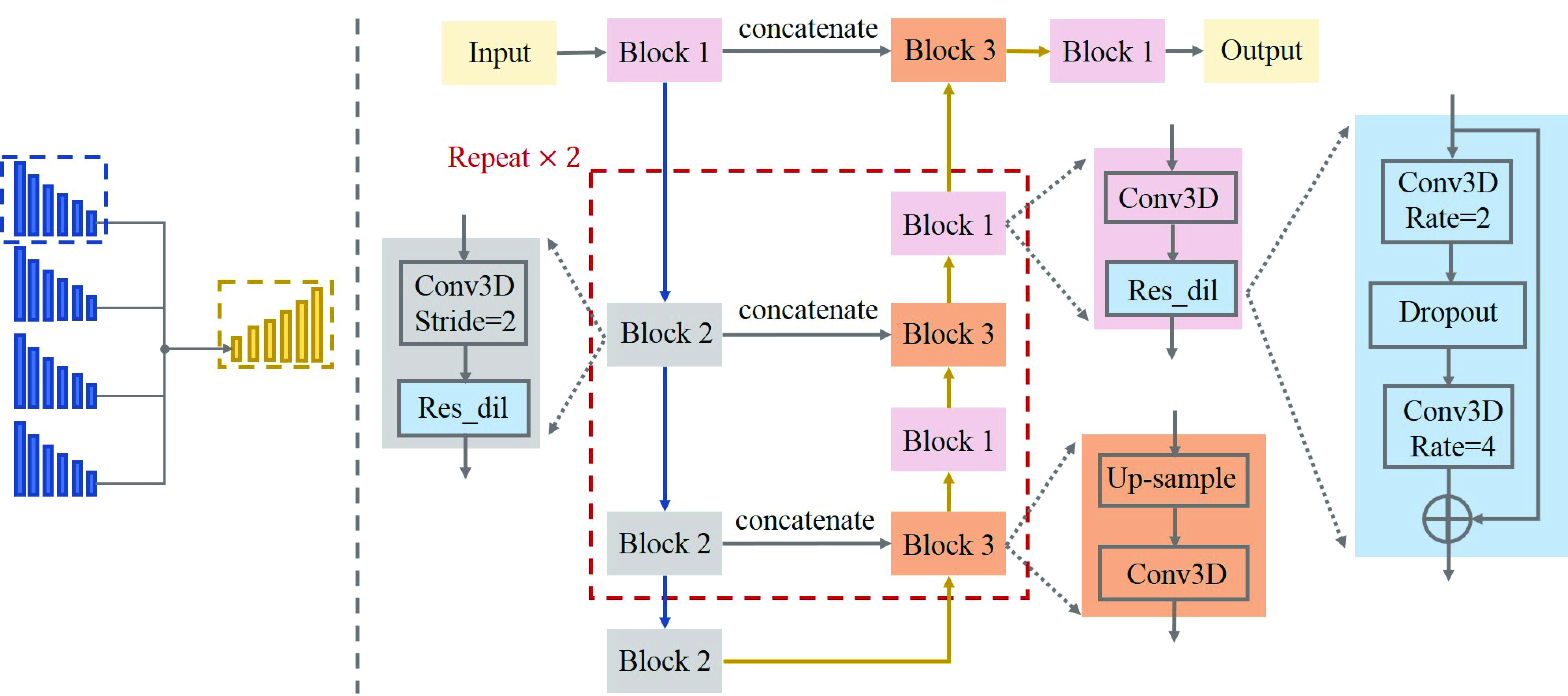}
\caption{Architecture of the Feature-enhanced Generator (FeG). Left: multi-encoder feature-enhanced generator shown in Figure \ref{fig2}; Right: we take one encoder and the decoder as an example. The left series of blocks connected by the blue arrows represent the encoder, and right series of blocks connected by the yellow arrows represent the decoder. In order to maintain the spatial information, we replace the pooling operation with the convolution block with stride=2. The res\_dil block used in both encoder and decoder is to increase the receptive field to capture more features, and the rate denotes the dilated rate.}
\label{fig3}
\end{figure*}

\begin{figure}[!t]
\centering
\subfloat[]{\includegraphics[width=1.5in]{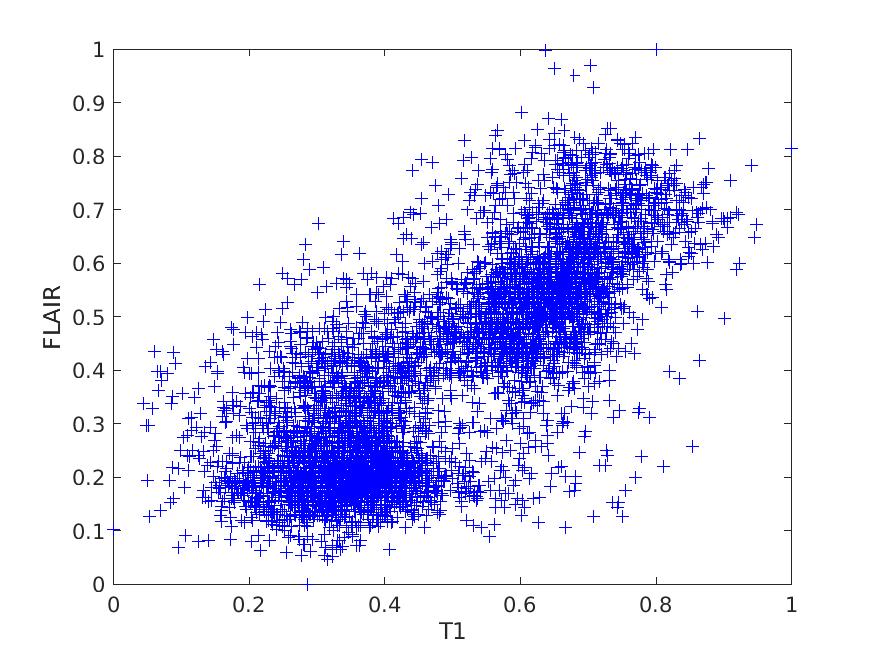}%
}
\hfil
\subfloat[]{\includegraphics[width=1.5in]{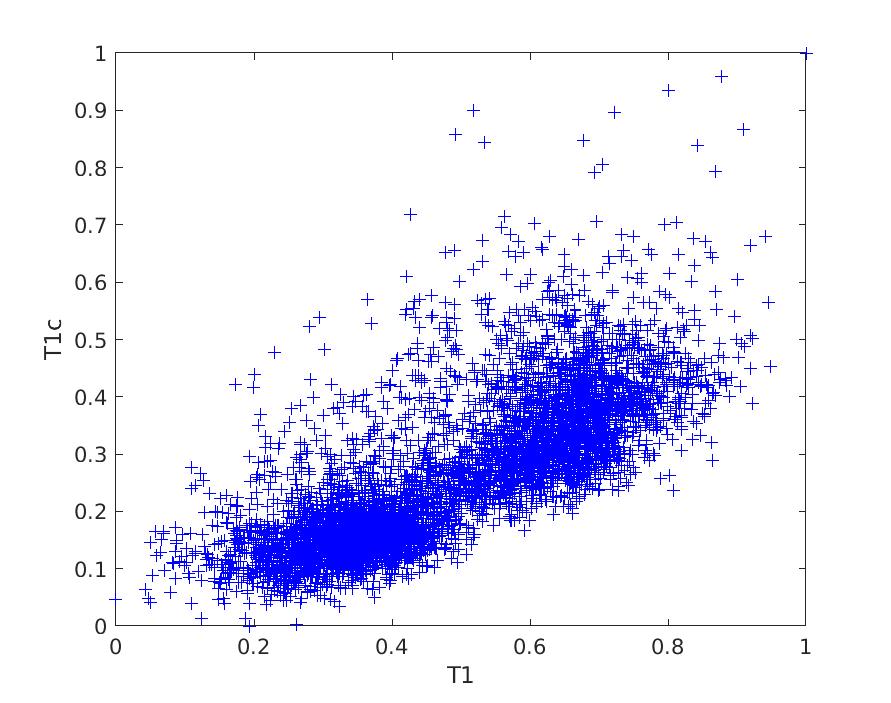}%
}
\hfil
\subfloat[]{\includegraphics[width=1.5in]{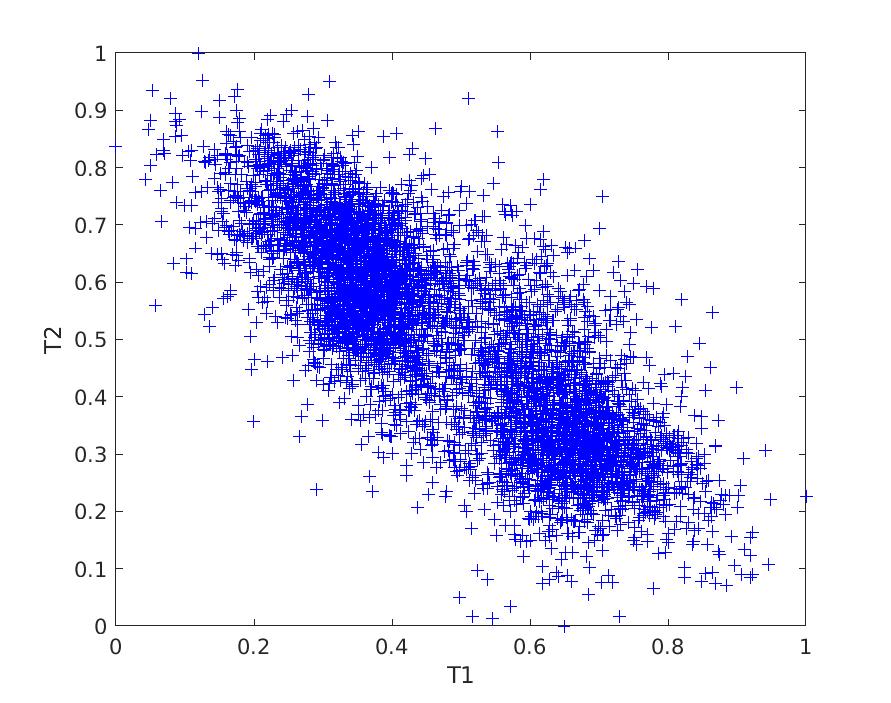}%
}
\hfil
\subfloat[]{\includegraphics[width=1.5in]{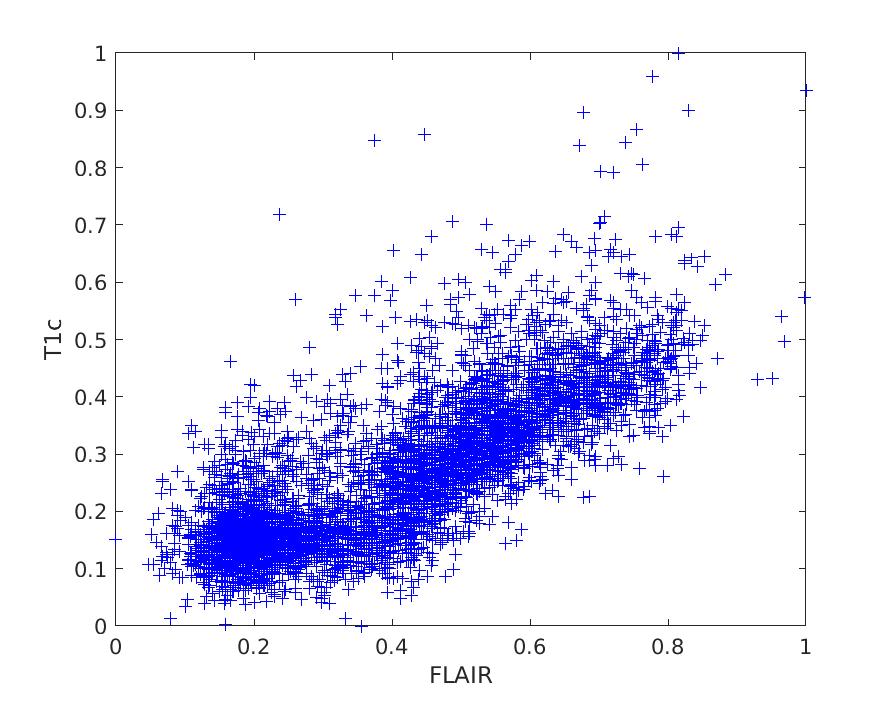}%
}
\hfil
\subfloat[]{\includegraphics[width=1.5in]{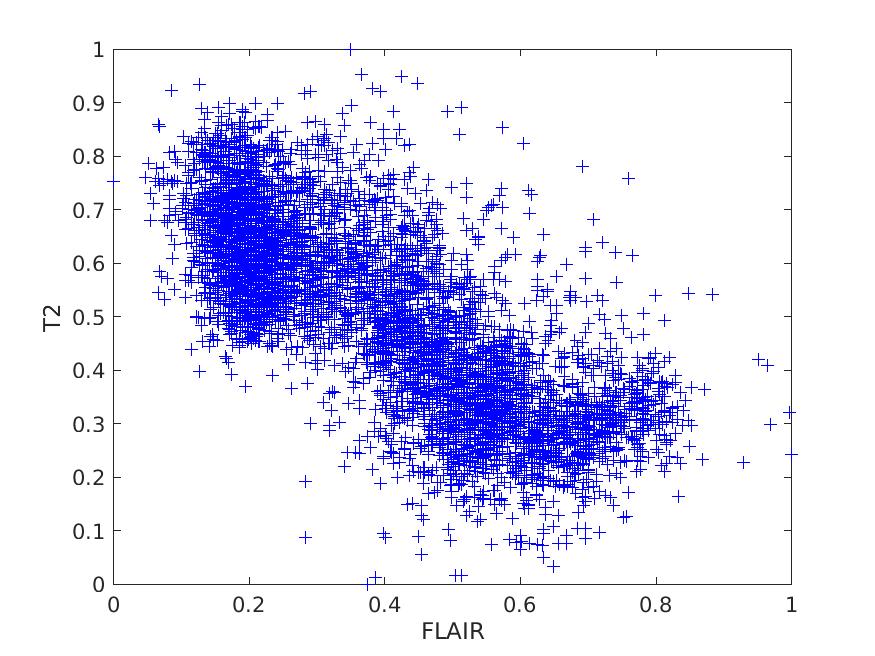}%
}
\hfil
\subfloat[]{\includegraphics[width=1.5in]{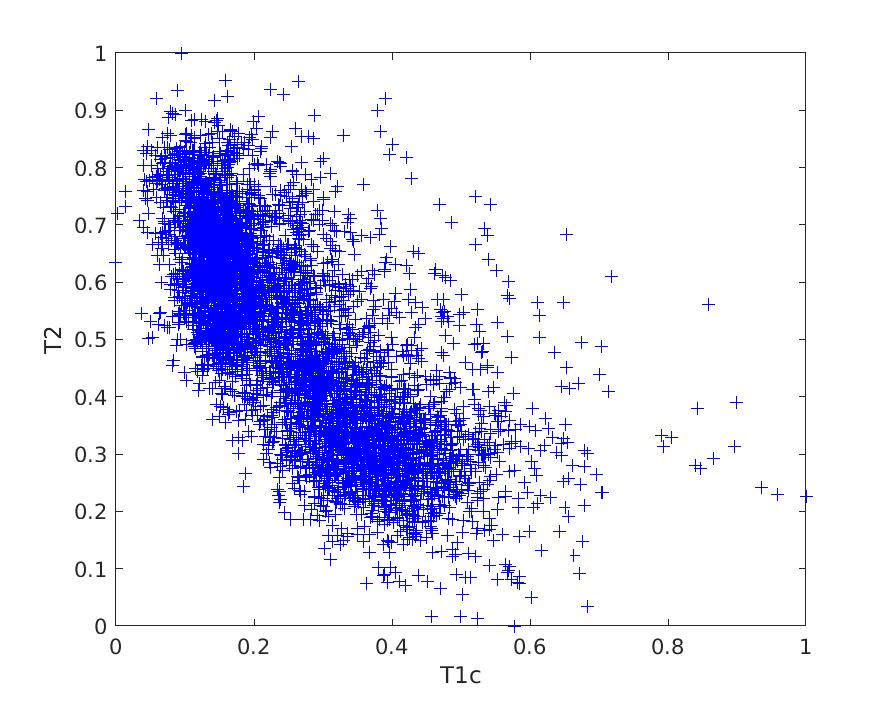}%
}
\caption{Joint intensity distributions of MR images: (a) T1-FLAIR, (b) T1-T1c, (c) T1-T2, (d) FLAIR-T1c, (e) FLAIR-T2, (f) T1c-T2. The intensity of the first modality is read on abscissa axis and that of the second modality on the ordinate axis.}
\label{fig4}
\end{figure}

Generative Adversarial Network (GAN) \cite{goodfellow2014generative} has demonstrated to be a promising approach for image synthesis. However, mode collapse, non-convergence, instability, and highly sensibility to hyper-parameters make it difficult for training. In addition, compared to a 2D image, it is more difficult to generate a 3D image volume. And our goal is to segment brain tumor using the available modalities, not to synthesize a perfect missing modality. Therefore, we designed a special Feature-enhanced Generator, called FeG, which used an average image of the missing modalities as the ground truth. Here, we use the average image as the ground truth for two reasons: (i) The averaging operation is simple to realize. And in this case, one decoder is sufficient to cope with any number of missing modalities. No additional decoders are required for each missing modality; (ii) Since the features from different modalities are correlated in one single patient. The averaging operation can extract the overall feature representations from multi-modalities, providing necessary feature representations for the following segmentation network. Specifically, we adopted a multi-encoder based network to generate the missing modalities. The independent encoders can not only learn the modality-specific feature representations but also can avoid the false-adaptation between modalities. The architecture is shown in Figure \ref{fig3}. Specifically, each encoder starts with block 1 in the first level, which consists of a 3D convolution layer and followed by a res\_dil block. Then, block 2 is applied in the following levels, which consists of a convolutional block with $stride = 2$ and a res\_dil block. The res\_dil block is inspired by the residual block and dilated convolution, it consists of a 3D dilated convolution layer ($rate =2$), a dropout layer ($rate=0.3$), and another 3D dilated convolution layer ($rate =4$). A residual connection is used in this block. It can increase the receptive field and help the network obtain more semantic features. The decoder begins with block 3, which consists of a up-sampling layer and a 3D convolution layer. Then the up-sampled features are concatenated with the features from the corresponding level of the encoders. Following the concatenation, block 1 is used to first adjust the number of features, and then enlarge the receptive field.

It is noted that, the generator and the segmentation network share the same encoders for the available modalities. There are three advantages of that: (i) The sharing operation can simplify the network architecture and reduce the training parameters. (ii) The feature-enhanced generator can learn the tumor related feature information from the segmentation network, making the interested regions obvious in the generated image volume. (iii) The segmentation network can utilize the generated feature-enhanced modality to improve the segmentation performance.

\subsection{Modelling the multi-source correlation}
\label{sec3.2}

\begin{figure}[!t]
\centering
\includegraphics[width=\columnwidth]{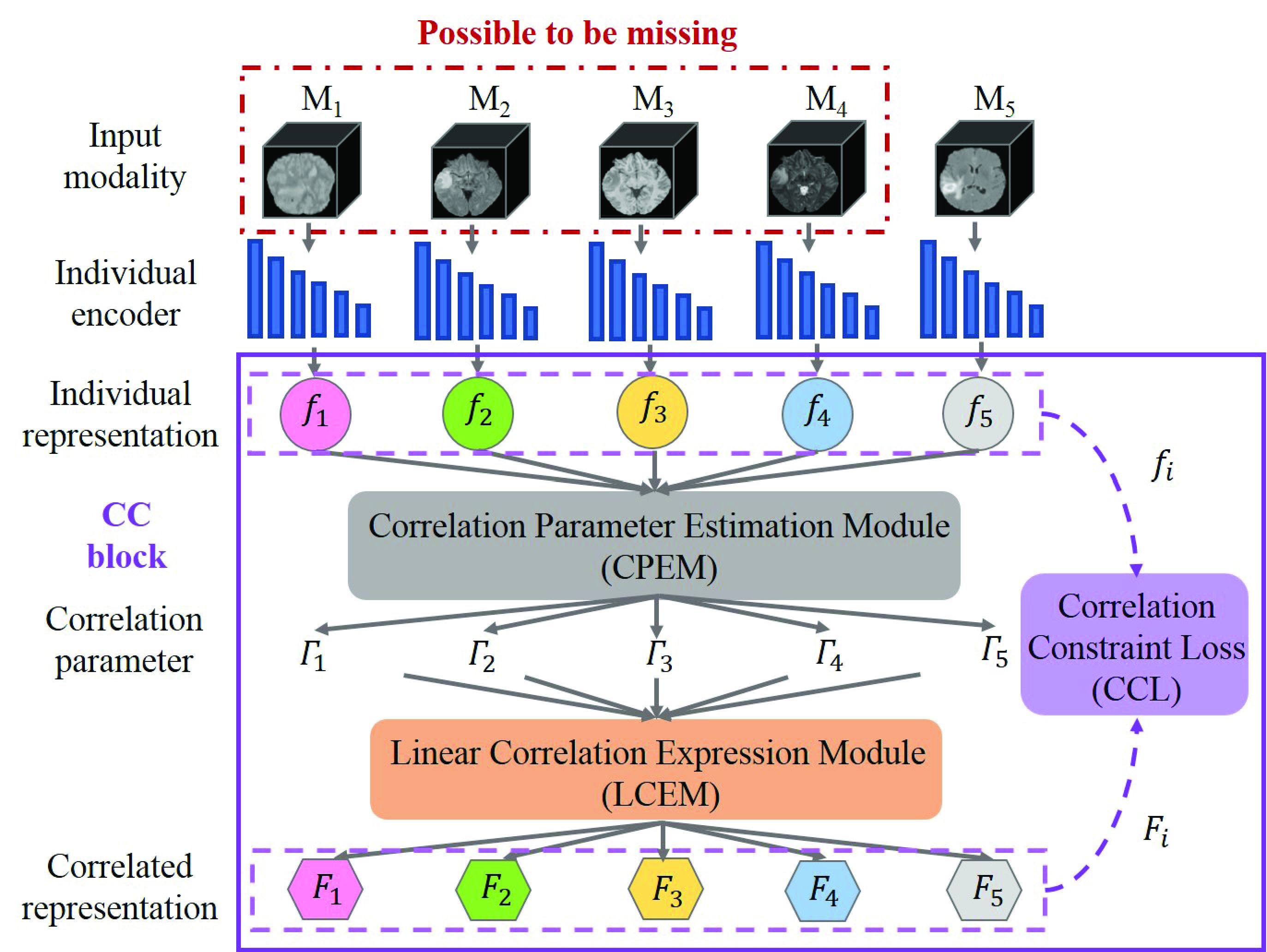}
\caption{Architecture of the Correlation Constraint (CC) block. $M_1$, $M_2$, $M_3$, $M_4$ are the input modalities, which are possible to be missing. $M_5$ is the generated average modality. CPEM first maps the individual representation $f_i$ to a set of independent parameters $\Gamma_i$, under these parameters, LCEM transforms all the individual representations to form correlated representation $F_i$. In addition, the KL based correlation constraint loss (CCL) is employed to guide the whole training process.}
\label{fig5}
\end{figure}

The proposed CC block is presented in Figure \ref{fig5}. It consists of three components: Correlation Parameter Estimation Module (CPEM), Linear Correlation Expression Module (LCEM) and Correlation Constraint Loss (CCL). After generating the feature-enhanced modality, a new complete modalities are obtained. Each input modality $\{X_i\}$, where $i=\{1,2,3,4,5\}$, is input to the independent encoder to learn the modality-specific feature representation $f_i(X_i|\theta_i)$, where $\theta_i$ denotes the parameters used in $ith$ encoder, such as the number of filters, the kernel size of filter and the rate of dropout. It is noted that only the generated feature-enhanced modality needs to train an encoder to get the independent feature representation, the other encoders are directly taken from the feature-enhanced generator. Following that, the CPEM is first used to learn the correlated weight parameters for each modality, which is a network with two fully connected networks. It maps the modality-specific feature representation $f_i(X_i|\theta_i)$ to a set of weight parameters $\Gamma_i =\{\alpha_i, \beta_i, \gamma_i, \delta_i\, \sigma_i\}$. Then, the LCEM is employed to produce the correlated feature representation $F_i(X_i|\theta_i)$ for each modality (Equation ~\ref{eq1}). According to our observation, we assume the correlation in this work is linear. However, the proposed LCEM can be generally integrated to any multi-source correlation problem, and the specific correlated expression will depend on the application. Finally, the CCL is introduced, which is a Kullback–Leibler divergence based loss function (Equation~\ref{eq2}). It can constrain the distributions between the estimated correlated feature representation and the original feature representation to be as close as possible. 

In conclusion, on the one hand, the correlation constraint block can help the feature-enhanced generator to synthesize the missing modality which should keep the multi-source correlation with the available modalities. On the other hand, it can guide the segmentation network to learn the correlated feature representation to improve the segmentation performance.

\begin{equation}
\begin{aligned}
    &F_i(X_i|\theta_i) = \alpha_i \odot f_j(X_{j}|\theta_{j})+\beta_i \odot f_k(X_{k}|\theta_{k})+ \gamma_i \odot f_l(X_{l}|\theta_{l})+\\
    & \delta_i \odot f_m(X_{m}|\theta_{m})+\sigma_i
    , (i \neq j \neq k \neq l \neq m)
\end{aligned}
\label{eq1}
\end{equation}

\noindent where $X$ is the input modality, $i$, $j$, $k$, $l$ and $m$ are the indexes of the modality, and $i, j, k, l, m=\{1,2,3,4,5\}$, $\theta$ is the network parameters, $f$ is the independent feature representation, $F$ is the correlated feature representation, $\alpha$, $\beta$, $\gamma$, $\delta$ and $\sigma$ are the correlation weight parameters.

\begin{equation}
	\textcolor{blue}{L_{cc}} = \sum_{i=1}^M {P(f_i)} log\frac{P(f_i)}{Q(g_i)}
	\label{chap5-eq2}
\end{equation}

\noindent where $M$ is the number of modality, $P(f_i)$ and $Q(g_i)$ are probability distributions of original feature representation and correlated feature representation of modality $i$, respectively.

\subsection{Brain tumor segmentation}
\label{sec3.3}
The proposed segmentation network is adopted from our previous work \cite{zhou2020multi}. The multiple encoders are used to extract the independent feature representations for each modality. To emphasize the most important features from different modalities, a fusion block based on attention mechanism \cite{hu2018squeeze}\cite{roy2018concurrent} is introduced in each level of the network. It allows to selectively emphasize feature representations along spatial-wise and modality-wise. Moreover, the deep supervision~\cite{isensee2017brain} is employed by integrating the segmentation results from different levels to form the final network output.

\subsection{The choices of loss function}
We develop a hybrid loss function for our network, which is defined as follow:
\begin{equation}
    \ L_{total}= L_{seg} + \lambda L_{gen} + \eta L_{cc}
\end{equation}
\noindent where $L_{seg}$, $L_{gen}$ and $L_{cc}$ are designed for the segmentation network, feature-enhanced generator and the correlation constraint block, respectively. $\lambda$ and $\eta$ are the trade-off parameters, where $\lambda=0.1$ and $\eta=0.1$ in our work. 

For segmentation, we use dice loss to calculate the accuracy of the prediction results.
\begin{equation}
    \ L_{seg}=1-2\frac{\sum_{i=1}^N\ p_{i} g_{i}} {\sum_{i=1}^N (p_{i} + g_{i})}
\end{equation}
\noindent where $N$ is the set of all examples, $p_{i}$ is the prediction result, $g_{i}$ is the ground-truth.

For the feature-enhanced generator, we use Structural Similarity Index Metic (SSIM) as the loss function. 

\begin{equation}
  L_{gen}=1-\sum_{i=1}^N\frac{({2\mu_{i\hat{y}}}{\mu_{iy}}+c_1)({2\sigma_{iy{\hat{y}}}}+c_2)}{{(\mu_{i\hat{y}}}^2+{\mu_{iy}^2}+c_1)({{\sigma_{i\hat{y}}}^{2}}+\sigma_{iy}^{2}+c_2)}
\end{equation}

\noindent where $N$ is the set of all examples, $y_i$ is the real image, $\hat{y_i}$ is the generated image, $\mu$ and $\sigma$ are the mean and standard deviation of the image, and $\sigma_{y\hat{y}}$ is the co-variance between the real image $y$ and the generated image $\hat{y}$,  $c_1$ and $c_2$ are to stabilize the division with weak denominator. 

\section{Experimental setup}
\label{sec4}

\subsection{Dataset and pre-processing}
\label{sec4.1}
To evaluate our proposed method, we used BraTS 2018 dataset \cite{bakas2018identifying}. It contains 285 cases with ground-truth, each case has four image modalities including T1, T1c, T2 and FLAIR. Following the challenge, four intra-tumor structures (edema, enhancing tumor, necrotic and non-enhancing tumor core) have been grouped into three mutually inclusive tumor regions: 
\noindent (a) The whole tumor region (WT), consisting of all tumor tissues. 
\noindent (b) The tumor core region (TC), consisting of the enhancing tumor, necrotic and non-enhancing tumor core. 
\noindent (c) The enhancing tumor region (ET). 

The provided data have been pre-processed by organisers: co-registered to the same anatomical template, interpolated to the same resolution ($1 mm^3$) and skull-stripped. The ground-truth have been manually labeled by experts. We did additional pre-processing with a standard procedure. To exploit the spatial contextual information of the image, we used 3D image, we cropped and resized them from $155 \times 240 \times240$ to $128 \times 128 \times128$. The N4ITK \cite{avants2009advanced} method is used to do the bias field correction for MRI data, and intensity normalization is applied to normalize each modality to a zero-mean, unit-variance space. 

\subsection{Implementation details}
\label{sec4.2}
The proposed network is implemented using Keras with a single Nvidia Tesla V100 (32G). The model is trained using Nadam optimizer, the initial learning rate is 0.0005, it will reduce with a factor 0.5 with patience of 5 epochs. To avoid over-fitting, early stopping is used if the validation loss is not improved over 10 epochs. We randomly split the dataset into 80\% training and 20\% testing. All the results are obtained by online evaluation platform\footnote{https://ipp.cbica.upenn.edu/}.

\subsection{Evaluation metrics}
\label{sec4.3}
To obtain quantitative measurements of the segmentation accuracy, we used two evaluation metrics: Dice Similarity Coefficient (DSC) and Hausdorff Distance (HD).

\noindent 1) DSC: It is used to calculate the overlap rate of prediction result and the ground-truth. The better predict result will have a larger value.

\begin{equation}
 DSC= \frac {2TP}{2TP+FP+FN}
\end{equation}

\noindent where $TP$ represents the number of true positive voxels, $FP$ the number of false positive voxels, and $FN$ the number of false negative voxels. 

\noindent 2) HD: It is computed between boundaries of the prediction result and the ground-truth, it is an indicator of the largest segmentation error. The better predict result will have a smaller value.



\begin{equation}
		HD = \max\{\max\limits_{s\in S}\min\limits_{r\in R} d(s,r), \max\limits_{r\in R}\min\limits_{s\in S} d(r,s)\}
\end{equation}
\noindent where $S$ and $R$ are the two sets of the surface points of the prediction and the real annotation, and $d$ is the Euclidean distance.

\section{Experiment results}
\label{sec5}

We carry out a series of comparative experiments to demonstrate the effectiveness of our proposed method. In Section \ref{5.1.1}, we illustrate the advantages of the proposed components. In Section \ref{5.1.2}, we compare our method with the state-of-the-art approaches. In Section \ref{5.2}, we conduct the qualitative experiment to further demonstrate that our proposed method can obtain the promising segmentation results.

\subsection{Quantitative analysis}
\label{sec5.1}
\subsubsection{Segmentation performance analysis on the proposed components}
\label{5.1.1}
We first carry out a series of ablation experiments to investigate the relative contribution of the proposed components, including the Feature-enhanced Generator (FeG) and the Correlation Constraint (CC) block. The comparison results are presented in Table~\ref{tab1} and Table~\ref{tab2}. The baseline denotes our method without FeG and CC block. From Table~\ref{tab1}, we can observe that the baseline can obtain 79.6, 67.1, 49.4 and 65.4 in the terms of Dice Score on whole tumor, tumor core, enhancing tumor and the average result across all the situations, respectively. The proposed feature-enhanced generator can indeed improve the segmentation results compared to the baseline, with an improvement of 2.3\%, 3.7\%, 6.3\% and 3.7\% in terms of Dice Score on whole tumor, tumor core, enhancing tumor and the average result across all the situations, respectively. We explain that the proposed feature-enhanced generator can compensate the missing data and help the network to improve the segmentation performance. 

It can also be observed that with the assistance of correlation constraint block, the segmentation results have significant improvements in all the cases. And statistically significant differences (an improvement of 15.5\% compared to the 'Baseline + FeG') can be observed when only FLAIR and T2 are available. The comparison results demonstrate that the multi-source correlation constraint block can guide the segmentation network to focus on the related target feature representations so as to improve the segmentation performance. 

The similar comparison results in terms of Hausdorff Distance can be observed in Table~\ref{tab2}. It is noticed that, the average Hausdorff Distance for 'Baseline', 'Baseline + FeG' and ours are 10.0, 8.7 and 7.1, respectively. We can observe an improvement of 13.0\% by the FeG and 18.4\% by the CC block, respectively, which indicates that the proposed additional generator and the correlation constrain block are really helpful for the segmentation. In conclusion, the comparison results demonstrate the effectiveness of the proposed component.

\begin{table*}[]
\centering
\caption{Comparison of different methods in terms of Dice Score on BraTS 2018 dataset, $\bullet$ denotes the present modality and $\circ$ denotes the missing one, bold results denotes the best score. WT, TC, ET denote whole tumor, tumor core and enhancing tumor, respectively. AVG denotes the average results on the three target regions, Average denotes the average results on one target region across all the situations.}
\label{tab1}
\resizebox{0.8\textwidth}{!}{%
\begin{tabular}{cccc|cccc|cccc|cccc}
\hline
\multicolumn{4}{c|}{Modality} & \multicolumn{4}{c|}{Baseline} & \multicolumn{4}{c|}{Baseline + FeG} & \multicolumn{4}{c}{Baseline + FeG + CC (Ours)} \\ \hline
F & T1 & T1c & T2 & WT & TC & ET & \cellcolor[HTML]{EFEFEF}AVG & WT & TC & ET & \cellcolor[HTML]{EFEFEF}AVG & WT & TC & ET & \cellcolor[HTML]{EFEFEF}AVG \\ \hline
$\circ$ & $\circ$ & $\circ$ & $\bullet$ & 76.5 & 50.3 & 21.4 & \cellcolor[HTML]{EFEFEF}49.4 & 78.5 & 52.6 & 25.5 & \cellcolor[HTML]{EFEFEF}52.2 & \textbf{80.4} & \textbf{59.5} & \textbf{35.2} & \cellcolor[HTML]{EFEFEF}\textbf{58.4} \\
$\circ$ & $\circ$ & $\bullet$ & $\circ$ & 65.6 & 78.1 & 70.5 & \cellcolor[HTML]{EFEFEF}71.4 & 68.4 & 79.6 & 71.6 & \cellcolor[HTML]{EFEFEF}73.2 & \textbf{72.0} & \textbf{83.1} & \textbf{75.0} & \cellcolor[HTML]{EFEFEF}\textbf{76.7} \\
$\circ$ & $\bullet$ & $\circ$ & $\circ$ & 66.2 & 39.0 & 17.5 & \cellcolor[HTML]{EFEFEF}40.9 & 69.8 & 47.5 & 16.6 & \cellcolor[HTML]{EFEFEF}44.6 & \textbf{74.9} & \textbf{55.5} & \textbf{29.2} & \cellcolor[HTML]{EFEFEF}\textbf{53.2} \\
$\bullet$ & $\circ$ & $\circ$ & $\circ$ & 84.2 & 50.0 & 16.3 & \cellcolor[HTML]{EFEFEF}50.2 & 85.3 & 53.6 & 25.8 & \cellcolor[HTML]{EFEFEF}54.9 & \textbf{85.9} & \textbf{64.6} & \textbf{39.5} & \cellcolor[HTML]{EFEFEF}\textbf{63.3} \\
$\circ$ & $\circ$ & $\bullet$ & $\bullet$ & 78.2 & 80.6 & 71.2 & \cellcolor[HTML]{EFEFEF}76.7 & 81.0 & 83.3 & 74.7 & \cellcolor[HTML]{EFEFEF}79.7 & \textbf{81.7} & \textbf{84.8} & \textbf{75.5} & \cellcolor[HTML]{EFEFEF}\textbf{80.7} \\
$\circ$ & $\bullet$ & $\bullet$ & $\circ$ & 70.0 & 79.0 & 71.0 & \cellcolor[HTML]{EFEFEF}73.3 & 74.3 & 81.4 & 73.9 & \cellcolor[HTML]{EFEFEF}76.5 & \textbf{76.2} & \textbf{84.2} & \textbf{75.8} & \cellcolor[HTML]{EFEFEF}\textbf{78.7} \\
$\bullet$ & $\bullet$ & $\circ$ & $\circ$ & 84.4 & 53.8 & 21.1 & \cellcolor[HTML]{EFEFEF}53.1 & 85.8 & 57.6 & 28.5 & \cellcolor[HTML]{EFEFEF}57.3 & \textbf{86.5} & \textbf{67.0} & \textbf{43.2} & \cellcolor[HTML]{EFEFEF}\textbf{65.6} \\
$\circ$ & $\bullet$ & $\circ$ & $\bullet$ & 77.9 & 51.5 & 25.6 & \cellcolor[HTML]{EFEFEF}51.7 & 80.8 & 54.1 & 27.8 & \cellcolor[HTML]{EFEFEF}54.2 & \textbf{83.4} & \textbf{62.6} & \textbf{38.0} & \cellcolor[HTML]{EFEFEF}\textbf{61.3} \\
$\bullet$ & $\circ$ & $\circ$ & $\bullet$ & 84.1 & 54.7 & 24.2 & \cellcolor[HTML]{EFEFEF}54.3 & 85.8 & 56.2 & 30.0 & \cellcolor[HTML]{EFEFEF}57.3 & \textbf{86.5} & \textbf{66.6} & \textbf{45.5} & \cellcolor[HTML]{EFEFEF}\textbf{66.2} \\
$\bullet$ & $\circ$ & $\bullet$ & $\circ$ & 84.7 & 80.8 & 72.4 & \cellcolor[HTML]{EFEFEF}79.3 & 85.5 & 84.2 & 75.9 & \cellcolor[HTML]{EFEFEF}81.9 & \textbf{86.2} & \textbf{85.0} & \textbf{77.1} & \cellcolor[HTML]{EFEFEF}\textbf{82.8} \\
$\bullet$ & $\bullet$ & $\bullet$ & $\circ$ & 85.7 & 82.9 & 75.6 & \cellcolor[HTML]{EFEFEF}81.4 & 85.8 & 84.2 & 76.6 & \cellcolor[HTML]{EFEFEF}82.2 & \textbf{86.6} & \textbf{85.6} & \textbf{77.2} & \cellcolor[HTML]{EFEFEF}\textbf{83.1} \\
$\bullet$ & $\bullet$ & $\circ$ & $\bullet$ & 85.5 & 58.4 & 32.0 & \cellcolor[HTML]{EFEFEF}58.6 & 86.2 & 58.6 & 32.2 & \cellcolor[HTML]{EFEFEF}59.0 & \textbf{86.8} & \textbf{68.0} & \textbf{45.6} & \cellcolor[HTML]{EFEFEF}\textbf{66.8} \\
$\bullet$ & $\circ$ & $\bullet$ & $\bullet$ & 85.9 & 84.3 & 75.4 & \cellcolor[HTML]{EFEFEF}81.9 & 85.9 & 85.0 & 76.5 & \cellcolor[HTML]{EFEFEF}82.5 & \textbf{86.4} & \textbf{86.0} & \textbf{76.9} & \cellcolor[HTML]{EFEFEF}\textbf{83.1} \\
$\circ$ & $\bullet$ & $\bullet$ & $\bullet$ & 80.9 & 81.9 & 74.3 & \cellcolor[HTML]{EFEFEF}79.0 & 82.1 & 82.5 & 75.0 & \cellcolor[HTML]{EFEFEF}79.9 & \textbf{82.9} & \textbf{85.2} & \textbf{76.2} & \cellcolor[HTML]{EFEFEF}\textbf{81.4} \\
$\bullet$ & $\bullet$ & $\bullet$ & $\bullet$ & 84.7 & 81.4 & 72.1 & \cellcolor[HTML]{EFEFEF}79.4 & 85.9 & 84.3 & 76.7 & \cellcolor[HTML]{EFEFEF}82.3 & \textbf{86.6} & \textbf{85.8} & \textbf{76.9} & \cellcolor[HTML]{EFEFEF}\textbf{83.1} \\ \hline
\multicolumn{4}{c|}{Average} & 79.6 & 67.1 & 49.4 & \cellcolor[HTML]{EFEFEF}65.4 & 81.4 & 69.6 & 52.5 & \cellcolor[HTML]{EFEFEF}67.8 & \textbf{82.9} & \textbf{74.9} & \textbf{59.1} & \cellcolor[HTML]{EFEFEF}\textbf{72.3} \\ \hline
\end{tabular}%
}
\end{table*}

\begin{table*}[]
\centering
\caption{Comparison of different methods in terms of Hausdorff Distance on BraTS 2018 dataset, $\bullet$ denotes the present modality and $\circ$ denotes the missing one, bold results denotes the best score. WT, TC, ET denote whole tumor, tumor core and enhancing tumor, respectively. AVG denotes the average results on the three target regions, Average denotes the average results on one target region across all the situations.}
\label{tab2}
\resizebox{0.8\textwidth}{!}{%
\begin{tabular}{cccc|cccc|cccc|cccc}
\hline
\multicolumn{4}{c|}{} & \multicolumn{4}{c|}{Baseline} & \multicolumn{4}{c|}{Baseline + FeG} & \multicolumn{4}{c}{Baseline + FeG + CC (Ours)} \\ \hline
F & T1 & T1c & T2 & WT & TC & ET & \cellcolor[HTML]{EFEFEF}AVG & WT & TC & ET & \cellcolor[HTML]{EFEFEF}AVG & WT & TC & ET & \cellcolor[HTML]{EFEFEF}AVG \\ \hline
$\circ$ & $\circ$ & $\circ$ & $\bullet$ & 13.3 & 16.1 & 15.1 & \cellcolor[HTML]{EFEFEF}14.8 & 10.6 & 15.0 & 14.7 & \cellcolor[HTML]{EFEFEF}13.4 & \textbf{9.0} & \textbf{12.2} & \textbf{11.9} & \cellcolor[HTML]{EFEFEF}\textbf{11.0} \\
$\circ$ & $\circ$ & $\bullet$ & $\circ$ & 15.9 & 10.5 & 8.0 & \cellcolor[HTML]{EFEFEF}11.5 & 12.1 & 8.9 & 6.2 & \cellcolor[HTML]{EFEFEF}9.1 & \textbf{10.8} & \textbf{6.5} & \textbf{4.5} & \cellcolor[HTML]{EFEFEF}\textbf{7.3} \\
$\circ$ & $\bullet$ & $\circ$ & $\circ$ & 15.7 & 23.9 & 22.5 & \cellcolor[HTML]{EFEFEF}20.7 & 14.3 & 18.3 & 18.2 & \cellcolor[HTML]{EFEFEF}16.9 & \textbf{10.0} & \textbf{15.2} & \textbf{13.4} & \cellcolor[HTML]{EFEFEF}\textbf{12.9} \\
$\bullet$ & $\circ$ & $\circ$ & $\circ$ & 5.9 & 13.5 & 13.3 & \cellcolor[HTML]{EFEFEF}10.9 & 6.2 & 13.3 & 12.2 & \cellcolor[HTML]{EFEFEF}10.6 & \textbf{5.4} & \textbf{10.9} & \textbf{9.7} & \cellcolor[HTML]{EFEFEF}\textbf{8.7} \\
$\circ$ & $\circ$ & $\bullet$ & $\bullet$ & 7.6 & 6.1 & 4.8 & \cellcolor[HTML]{EFEFEF}6.2 & 7.0 & 5.3 & 4.0 & \cellcolor[HTML]{EFEFEF}5.4 & \textbf{6.6} & \textbf{5.1} & \textbf{3.9} & \cellcolor[HTML]{EFEFEF}\textbf{5.2} \\
$\circ$ & $\bullet$ & $\bullet$ & $\circ$ & 14.0 & 11.5 & 9.9 & \cellcolor[HTML]{EFEFEF}11.8 & 9.3 & 7.9 & 5.0 & \cellcolor[HTML]{EFEFEF}7.4 & \textbf{8.7} & \textbf{5.4} & \textbf{4.2} & \cellcolor[HTML]{EFEFEF}\textbf{6.1} \\
$\bullet$ & $\bullet$ & $\circ$ & $\circ$ & 6.3 & 12.7 & 12.3 & \cellcolor[HTML]{EFEFEF}10.4 & 5.7 & 12.8 & 12.0 & \cellcolor[HTML]{EFEFEF}10.2 & \textbf{5.2} & \textbf{10.4} & \textbf{9.4} & \cellcolor[HTML]{EFEFEF}\textbf{8.3} \\
$\circ$ & $\bullet$ & $\circ$ & $\bullet$ & 10.7 & 17.3 & 16.6 & \cellcolor[HTML]{EFEFEF}14.9 & 8.5 & 14.7 & 12.8 & \cellcolor[HTML]{EFEFEF}12.0 & \textbf{6.6} & \textbf{10.3} & \textbf{11.3} & \cellcolor[HTML]{EFEFEF}\textbf{9.4} \\
$\bullet$ & $\circ$ & $\circ$ & $\bullet$ & 5.9 & 12.3 & 11.7 & \cellcolor[HTML]{EFEFEF}10.0 & 5.7 & 12.7 & 12.0 & \cellcolor[HTML]{EFEFEF}10.1 & \textbf{5.2} & \textbf{10.2} & \textbf{9.2} & \cellcolor[HTML]{EFEFEF}\textbf{8.2} \\
$\bullet$ & $\circ$ & $\bullet$ & $\circ$ & 5.8 & 5.0 & 3.8 & \cellcolor[HTML]{EFEFEF}4.9 & 5.4 & 5.5 & 4.3 & \cellcolor[HTML]{EFEFEF}5.1 & \textbf{5.0} & \textbf{4.6} & \textbf{3.3} & \cellcolor[HTML]{EFEFEF}\textbf{4.3} \\
$\bullet$ & $\bullet$ & $\bullet$ & $\circ$ & 6.3 & 5.4 & 3.5 & \cellcolor[HTML]{EFEFEF}5.1 & 5.3 & 5.6 & 4.1 & \cellcolor[HTML]{EFEFEF}5.0 & \textbf{5.0} & \textbf{4.2} & \textbf{3.0} & \cellcolor[HTML]{EFEFEF}\textbf{4.1} \\
$\bullet$ & $\bullet$ & $\circ$ & $\bullet$ & 6.4 & 12.8 & 11.4 & \cellcolor[HTML]{EFEFEF}10.2 & 5.4 & 12.6 & 11.3 & \cellcolor[HTML]{EFEFEF}9.8 & \textbf{5.2} & \textbf{9.0} & \textbf{8.9} & \cellcolor[HTML]{EFEFEF}\textbf{7.7} \\
$\bullet$ & $\circ$ & $\bullet$ & $\bullet$ & 6.9 & 4.6 & 3.6 & \cellcolor[HTML]{EFEFEF}5.0 & 5.5 & 5.5 & 4.2 & \cellcolor[HTML]{EFEFEF}5.1 & \textbf{5.2} & \textbf{4.1} & \textbf{3.2} & \cellcolor[HTML]{EFEFEF}\textbf{4.2} \\
$\circ$ & $\bullet$ & $\bullet$ & $\bullet$ & 9.7 & 8.1 & 6.8 & \cellcolor[HTML]{EFEFEF}8.2 & \textbf{6.3} & 5.4 & 3.9 & \cellcolor[HTML]{EFEFEF}5.2 & 6.5 & \textbf{4.9} & \textbf{3.5} & \cellcolor[HTML]{EFEFEF}\textbf{5.0} \\
$\bullet$ & $\bullet$ & $\bullet$ & $\bullet$ & \textbf{4.7} & 5.7 & 4.2 & \cellcolor[HTML]{EFEFEF}4.9 & 5.3 & 5.6 & 4.0 & \cellcolor[HTML]{EFEFEF}5.0 & 5.2 & \textbf{4.2} & \textbf{3.0} & \cellcolor[HTML]{EFEFEF}\textbf{4.1} \\ \hline
\multicolumn{4}{c|}{Average} & 9.0 & 11.0 & 9.8 & \cellcolor[HTML]{EFEFEF}10.0 & 7.5 & 9.9 & 8.6 & \cellcolor[HTML]{EFEFEF}8.7 & \textbf{6.6} & \textbf{7.8} & \textbf{6.8} & \cellcolor[HTML]{EFEFEF}\textbf{7.1} \\ \hline
\end{tabular}%
}
\end{table*}

\subsubsection{Comparison with the state-of-the-art methods}
\label{5.1.2}
The main contribution of our method is using the correlation constraint block to discover the latent multi-source correlation between modalities and make the model robust in the case of missing modalities. To demonstrate the advantages of our proposed method on missing modalities, we compare it with the state-of-the-art methods, which have been introduced in the related work. The comparison results are illustrated in Table~\ref{tab3}. Since the method HeMIS didn't publish the available code, the reported results on HeMIS and U-HeMIS \cite{havaei2016hemis} are taken from the work in \cite{dorent2019hetero}. For all the tumor regions, our method achieves the best results in most of the cases. Compared to the current state-of-the-art method \cite{dorent2019hetero}, our method can achieve the average Dice Score of 82.9, 74.9 and 59.1 across all the situations, which outperforms the best method by 3.5\%, 17\% and 18.2\%. We explain that the proposed feature-enhanced generator compensates the missing modality with the correlation constraint block. It indicated the importance of the complete dataset, since it can provide the full information for the network to learn the effective features for segmentation. We also compared with the method trained on missing-one modality \cite{zhu2021brain}. Our method can obtain much better results with a large margin, we can achieve the improvement of 6.7\%, 4.4\%, 7.4\% and 12.6\% in the terms of average Dice Score when T2, T1c, T1, Flair is missing, respectively. And for the full modalities, we can also outperform it by 6.3\% in the terms of average Dice Score.

From the comparison results, we can also obtain another observation. Missing FLAIR modality leads to a sharp decreasing on dice score for all the regions, since FLAIR is the principle modality for showing whole tumor. While missing T1c modality would have a severe decreasing on dice score for both tumor core and enhancing tumor, since T1c is the principle modality for showing tumor core and enhancing tumor regions. Missing T1 and T2 modalities would have a slight decreasing on dice score for all the regions. Furthermore, when only one modality is available, T1c modality can achieve the promising results. When two modalities are available, 'FLAIR + T1c' are the best combination, indicating that the importance of FLAIR and T1c for MR brain tumor segmentation. Our method can give better results in these cases than other methods.

\begin{table*}[]
\centering
\caption{Comparison of different methods in terms of Dice Score on BraTS 2018 dataset, $\bullet$ denotes the present modality and $\circ$ denotes the missing one, bold results denotes the best score. WT, TC, ET denote whole tumor, tumor core and enhancing tumor, respectively. AVG denotes the average results on the three target regions, Average denotes the average results on one target region across all the situations.}
\label{tab3}
\resizebox{\textwidth}{!}{%
\begin{tabular}{cccc|cccc|cccc|cccc|cccc|cccc|cccc}
\hline
\multicolumn{4}{c|}{Modality} & \multicolumn{4}{c|}{HeMIS \cite{havaei2016hemis}} & \multicolumn{4}{c|}{U-HeMIS \cite{havaei2016hemis}} & \multicolumn{4}{c|}{URN \cite{lau2019unified}} & \multicolumn{4}{c|}{U-HVED \cite{dorent2019hetero}} & \multicolumn{4}{c|}{\cite{zhu2021brain}} & \multicolumn{4}{c}{Ours} \\ \hline

F & T1 & T1c & T2 & WT & TC & ET & \cellcolor[HTML]{EFEFEF}AVG & WT & TC & ET & \cellcolor[HTML]{EFEFEF}AVG & WT & TC & ET & \cellcolor[HTML]{EFEFEF}AVG & WT & TC & ET & \cellcolor[HTML]{EFEFEF}AVG & WT & TC & ET & \cellcolor[HTML]{EFEFEF}AVG & WT & TC & ET & \cellcolor[HTML]{EFEFEF}AVG\\ \hline

$\circ$ & $\circ$ & $\circ$ & $\bullet$ & 38.6 & 19.5 & 0.0 & \cellcolor[HTML]{EFEFEF}19.4 & 79.2 & 50.0 & 23.3 & \cellcolor[HTML]{EFEFEF}50.8 & 77.5 & 43.6 & 20.3 & \cellcolor[HTML]{EFEFEF}47.1 & \textbf{80.9} & 54.1 & 30.8 & \cellcolor[HTML]{EFEFEF}55.3 &-&-&-& \cellcolor[HTML]{EFEFEF}-& 80.4 & \textbf{59.5} & \textbf{35.2} & \cellcolor[HTML]{EFEFEF}\textbf{58.4} \\

$\circ$ & $\circ$ & $\bullet$ & $\circ$ & 2.6 & 6.5 & 11.1 & \cellcolor[HTML]{EFEFEF}6.7 & 58.5 & 58.5 & 60.8 & \cellcolor[HTML]{EFEFEF}59.3 & 62.2 & 58.5 & 55.8 & \cellcolor[HTML]{EFEFEF}58.8 & 62.4 & 66.7 & 65.5 & \cellcolor[HTML]{EFEFEF}64.9 & -&-&-& \cellcolor[HTML]{EFEFEF}-& \textbf{72.0} & \textbf{83.1} & \textbf{75.0} & \cellcolor[HTML]{EFEFEF}\textbf{76.7} \\

$\circ$ & $\bullet$ & $\circ$ & $\circ$ & 0.0 & 0.0 & 0.0 & \cellcolor[HTML]{EFEFEF}0.0 & 54.3 & 37.9 & 12.4 & \cellcolor[HTML]{EFEFEF}34.9 & 50.4 & 34.2 & 19.1 & \cellcolor[HTML]{EFEFEF}34.6 & 52.4 & 37.2 & 13.7 & \cellcolor[HTML]{EFEFEF}34.4 &-&-&-& \cellcolor[HTML]{EFEFEF}-& \textbf{74.9} & \textbf{55.5} & \textbf{29.2} & \cellcolor[HTML]{EFEFEF}\textbf{53.2} \\

$\bullet$ & $\circ$ & $\circ$ & $\circ$ & 55.2 & 16.2 & 6.6 & \cellcolor[HTML]{EFEFEF}26.0 & 79.9 & 49.8 & 24.9 & \cellcolor[HTML]{EFEFEF}51.5 & 84.8 & 50.4 & 23.6 & \cellcolor[HTML]{EFEFEF}52.9 & 82.1 & 50.4 & 24.8 & \cellcolor[HTML]{EFEFEF}52.4 & -&-&-& \cellcolor[HTML]{EFEFEF}-&\textbf{85.9} & \textbf{64.6} & \textbf{39.5} & \cellcolor[HTML]{EFEFEF}\textbf{63.3} \\

$\circ$ & $\circ$ & $\bullet$ & $\bullet$ & 48.2 & 45.8 & 55.8 & \cellcolor[HTML]{EFEFEF}49.9 & 81.0 & 69.1 & 68.6 & \cellcolor[HTML]{EFEFEF}72.9 & 80.3 & 68.9 & 67.6 & \cellcolor[HTML]{EFEFEF}72.3 & \textbf{82.7} & 73.7 & 70.2 & \cellcolor[HTML]{EFEFEF}75.5 & -&-&-& \cellcolor[HTML]{EFEFEF}-&81.7 & \textbf{84.8} & \textbf{75.5} & \cellcolor[HTML]{EFEFEF}\textbf{80.7} \\

$\circ$ & $\bullet$ & $\bullet$ & $\circ$ & 15.4 & 30.4 & 42.6 & \cellcolor[HTML]{EFEFEF}29.5 & 63.8 & 64.0 & 65.3 & \cellcolor[HTML]{EFEFEF}64.4 & 69.8 & 65.9 & 66.5 & \cellcolor[HTML]{EFEFEF}67.4 & 66.8 & 69.7 & 67.0 & \cellcolor[HTML]{EFEFEF}67.8 & -&-&-& \cellcolor[HTML]{EFEFEF}-&\textbf{76.2} & \textbf{84.2} & \textbf{75.8} & \cellcolor[HTML]{EFEFEF}\textbf{78.7} \\

$\bullet$ & $\bullet$ & $\circ$ & $\circ$ & 71.1 & 11.9 & 1.2 & \cellcolor[HTML]{EFEFEF}28.1 & 83.9 & 56.7 & 29.0 & \cellcolor[HTML]{EFEFEF}56.5 & 85.5 & 52.6 & 25.3 & \cellcolor[HTML]{EFEFEF}54.5 & 84.3 & 55.3 & 24.2 & \cellcolor[HTML]{EFEFEF}54.6 & -&-&-& \cellcolor[HTML]{EFEFEF}-&\textbf{86.5} & \textbf{67.0} & \textbf{43.2} & \cellcolor[HTML]{EFEFEF}\textbf{65.6} \\

$\circ$ & $\bullet$ & $\circ$ & $\bullet$ & 47.3 & 17.2 & 0.6 & \cellcolor[HTML]{EFEFEF}21.7 & 80.8 & 53.4 & 28.3 & \cellcolor[HTML]{EFEFEF}54.2 & 80.8 & 48.6 & 25.2 & \cellcolor[HTML]{EFEFEF}51.5 & 82.2 & 57.2 & 30.7 & \cellcolor[HTML]{EFEFEF}56.7 & -&-&-& \cellcolor[HTML]{EFEFEF}-&\textbf{83.4} & \textbf{62.6} & \textbf{38.0} & \cellcolor[HTML]{EFEFEF}\textbf{61.3} \\

$\bullet$ & $\circ$ & $\circ$ & $\bullet$ & 74.8 & 17.7 & 0.8 & \cellcolor[HTML]{EFEFEF}31.1 & 86.0 & 58.7 & 28.0 & \cellcolor[HTML]{EFEFEF}57.6 & 86.3 & 50.7 & 25.2 & \cellcolor[HTML]{EFEFEF}54.1 & \textbf{87.5} & 59.7 & 34.6 & \cellcolor[HTML]{EFEFEF}60.6 & -&-&-& \cellcolor[HTML]{EFEFEF}-&86.5 & \textbf{66.6} & \textbf{45.5} & \cellcolor[HTML]{EFEFEF}\textbf{66.2} \\

$\bullet$ & $\circ$ & $\bullet$ & $\circ$ & 68.4 & 41.4 & 53.8 & \cellcolor[HTML]{EFEFEF}54.5 & 83.3 & 67.6 & 68.0 & \cellcolor[HTML]{EFEFEF}73.0 & 85.8 & 72.5 & 70.4 & \cellcolor[HTML]{EFEFEF}76.2 & 85.8 & 72.9 & 70.3 & \cellcolor[HTML]{EFEFEF}76.2 & -&-&-& \cellcolor[HTML]{EFEFEF}-&\textbf{86.2} & \textbf{85.0} & \textbf{77.1} & \cellcolor[HTML]{EFEFEF}\textbf{82.8} \\

$\bullet$ & $\bullet$ & $\bullet$ & $\circ$ & 70.2 & 48.8 & 60.9 & \cellcolor[HTML]{EFEFEF}60.0 & 85.1 & 70.7 & 69.9 & \cellcolor[HTML]{EFEFEF}75.2 & 85.6 & 72.0 & 71.0 & \cellcolor[HTML]{EFEFEF}76.2 & 86.2 & 74.2 & 71.1 & \cellcolor[HTML]{EFEFEF}77.2 & 86.1&78.2&69.3& \cellcolor[HTML]{EFEFEF}77.9&\textbf{86.6} & \textbf{85.6} & \textbf{77.2} & \cellcolor[HTML]{EFEFEF}\textbf{83.1} \\

$\bullet$ & $\bullet$ & $\circ$ & $\bullet$ & 75.2 & 18.7 & 1.0 & \cellcolor[HTML]{EFEFEF}31.6 & 87.0 & 61.0 & 33.4 & \cellcolor[HTML]{EFEFEF}60.5 & 86.1 & 52.5 & 25.8 & \cellcolor[HTML]{EFEFEF}54.8 & \textbf{88.0} & 61.5 & 34.1 & \cellcolor[HTML]{EFEFEF}61.2 &87.6&62.6&41.7& \cellcolor[HTML]{EFEFEF}64.0& 86.8 & \textbf{68.0} & \textbf{45.6} & \cellcolor[HTML]{EFEFEF}\textbf{66.8} \\

$\bullet$ & $\circ$ & $\bullet$ & $\bullet$ & 75.6 & 54.9 & 60.5 & \cellcolor[HTML]{EFEFEF}63.7 & 87.0 & 72.2 & 69.7 & \cellcolor[HTML]{EFEFEF}76.3 & 86.5 & 72.2 & 69.8 & \cellcolor[HTML]{EFEFEF}76.2 & \textbf{88.6} & 75.6 & 71.2 & \cellcolor[HTML]{EFEFEF}78.5 & 87.1&77.8&67.4& \cellcolor[HTML]{EFEFEF}77.4&86.4 & \textbf{86.0} & \textbf{76.9} & \cellcolor[HTML]{EFEFEF}\textbf{83.1} \\

$\circ$ & $\bullet$ & $\bullet$ & $\bullet$ & 44.2 & 46.6 & 55.1 & \cellcolor[HTML]{EFEFEF}48.6 & 82.1 & 70.7 & 69.7 & \cellcolor[HTML]{EFEFEF}74.2 & 81.1 & 69.5 & 68.5 & \cellcolor[HTML]{EFEFEF}73.0 & \textbf{83.3} & 75.3 & 71.1 & \cellcolor[HTML]{EFEFEF}76.6 & 75.6&76.0&65.3& \cellcolor[HTML]{EFEFEF}72.3&82.9 & \textbf{85.2} & \textbf{76.2} & \cellcolor[HTML]{EFEFEF}\textbf{81.4} \\

$\bullet$ & $\bullet$ & $\bullet$ & $\bullet$ & 73.8 & 55.3 & 61.1 & \cellcolor[HTML]{EFEFEF}63.4 & 87.6 & 73.4 & 70.8 & \cellcolor[HTML]{EFEFEF}77.3 & 86.3 & 71.8 & 69.9 & \cellcolor[HTML]{EFEFEF}76.0 & \textbf{88.8} & 76.4 & 71.7 & \cellcolor[HTML]{EFEFEF}79.0 & 88.3&77.7&68.5& \cellcolor[HTML]{EFEFEF}78.2&86.6 & \textbf{85.8} & \textbf{76.9} & \cellcolor[HTML]{EFEFEF}\textbf{83.1} \\ 

\hline
\multicolumn{4}{c|}{Average} & 50.7 & 28.7 & 27.4 & \cellcolor[HTML]{EFEFEF}28.1 & 78.6 & 59.7 & 48.1 & \cellcolor[HTML]{EFEFEF}62.1 & 79.3 & 58.9 & 46.9 & \cellcolor[HTML]{EFEFEF}61.7 & 80.1 & 64.0 & 50.0 & \cellcolor[HTML]{EFEFEF}64.7 &-&-&-& \cellcolor[HTML]{EFEFEF}-& \textbf{82.9} & \textbf{74.9} & \textbf{59.1} & \cellcolor[HTML]{EFEFEF}\textbf{72.3} \\ \hline
\end{tabular}%
}
\end{table*}

\subsection{Qualitative analysis}
\label{5.2}

\begin{figure}[!t]
\centering
\includegraphics[width=\columnwidth]{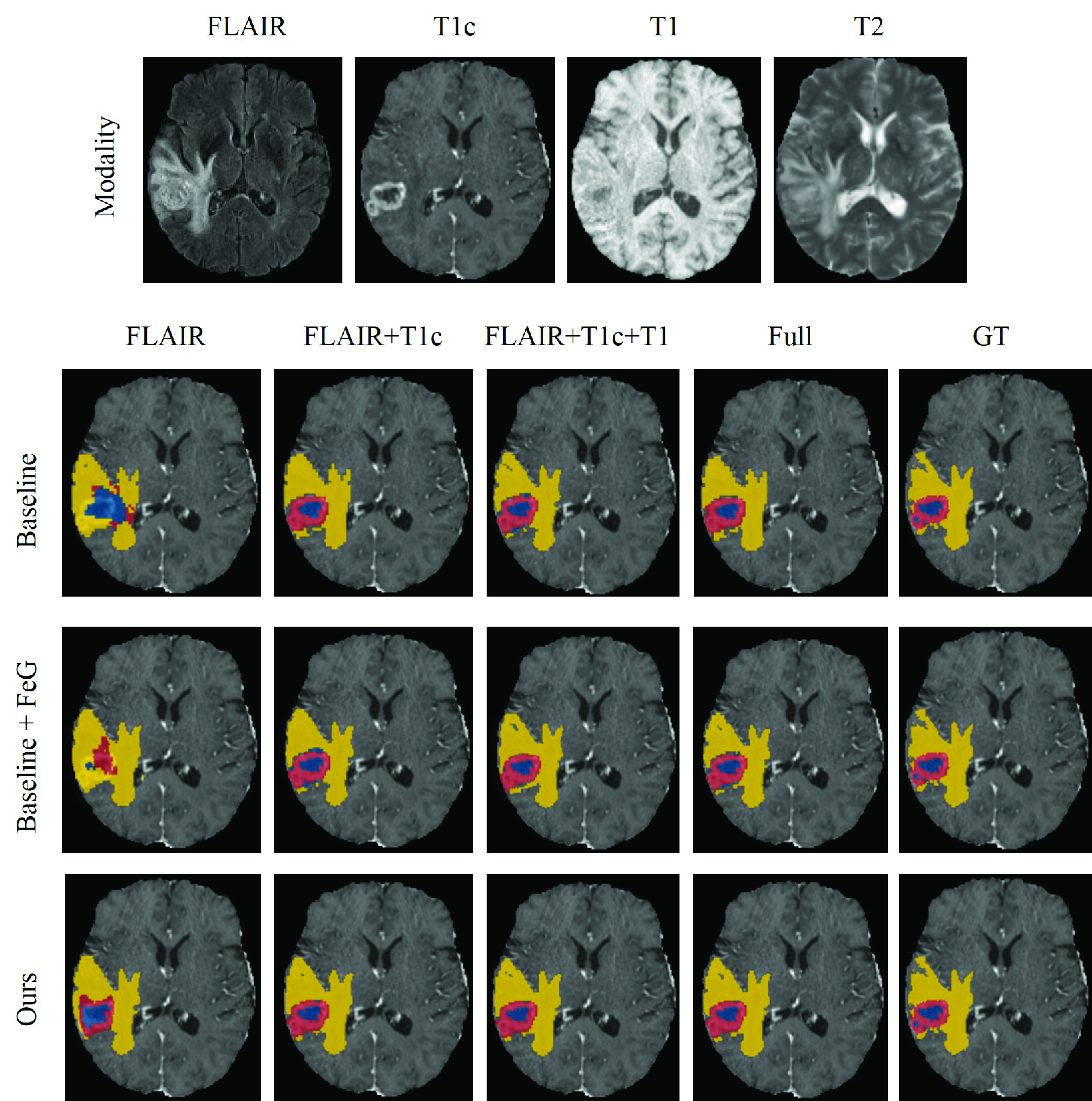}
\caption{Segmentation results of our proposed methods. On the top, the first row shows the four MR modalities, FLAIR, T1c, T1 and T2. On the bottom, the three rows show the segmentation results of different methods. The first four columns show the different missing modalities situations. The last column shows the ground-truth segmentation. Net\&ncr is shown in blue, edema in yellow and enhancing tumor in red. Net refers non-enhancing tumor and ncr necrotic tumor.}
\label{fig6}
\end{figure} 

\begin{figure}[!t]
\centering
\includegraphics[width=\columnwidth]{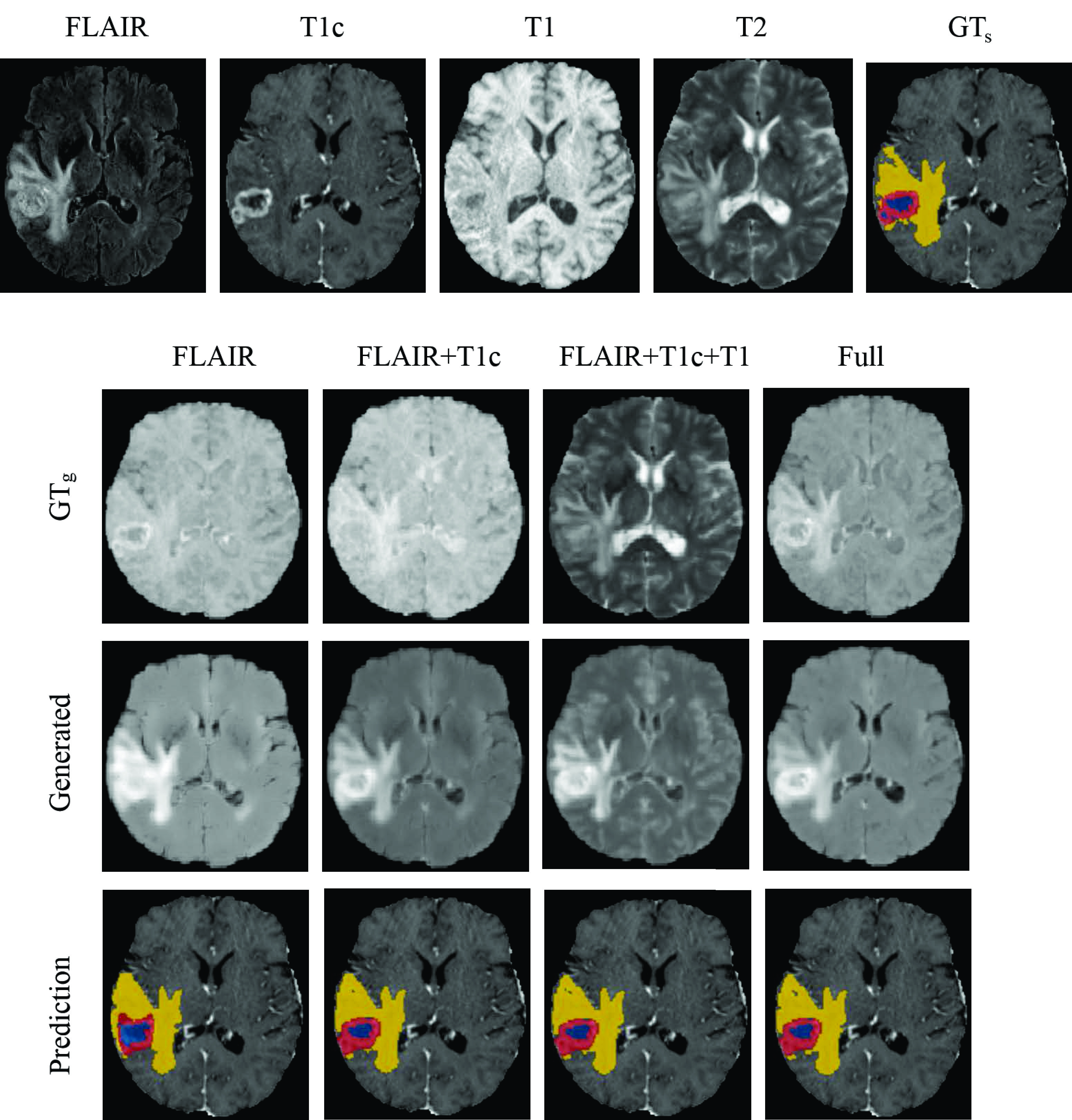}
\caption{Generation and segmentation results of our proposed methods. On the top, the first row shows the four MR modalities (FLAIR, T1c, T1 and T2) and the segmentation ground-truth. On the bottom, the four columns show the different missing modalities situations. The first row shows the ground-truth average modality. The second row shows the corresponding generated average modality. The last row shows the segmentation results of our method. Net\&ncr is shown in blue, edema in yellow and enhancing tumor in red. Net refers non-enhancing tumor and ncr necrotic tumor.}
\label{fig7}
\end{figure} 

To further demonstrate the performance of our proposed method, we first randomly select an example on BraTS 2018 dataset and visualize the results in Figure \ref{fig6} and Figure \ref{fig7}. In Figure \ref{fig6}, the first row shows the four MR modalities. The last three rows show the segmentation results. Here, 'Ours' denotes the method 'Baseline + FeG + CC'. From Figure \ref{fig6}, we can observe that in each column, with the help of our proposed components, the segmentation results can be gradually improved. Especially when only FLAIR is available, the baseline method produces many false predictions on the tumor core region. When the FeG is applied, these false predictions are corrected gradually. This improvement is mainly attributed to the new generated average modality, it can provide more rich information for the network to achieve more useful feature learning. Finally, with the help of the correlation constrain, we can obtain the best segmentation results. The reason for this improvement is that the effective feature representation guided by the correlation constrain can help the network to achieve the better performance. In each row, with the increase of the number of the modalities, the segmentation results become much better. It indicated that the full data is important to improve the segmentation accuracy. We can also find that with FLAIR, we can obtain a promising edema region. And when T1c is added, the tumor core region is well segmented. The T1 and T2 can help to refine the final results. 

To demonstrate the effectiveness of the proposed FeG, we also visualize the generated results in Figure \ref{fig7}. The first row presents the complete four MR modalities and the ground-truth. The second row shows the real average modality. The third row shows the generated average modality. The last row presents the corresponding segmentation results. We can observe that the target tumor regions are really obvious in the generated image. We explain that since the feature-enhanced generator shared the same encoders with the segmentation part, it makes the generator able to learn the related segmentation features. In addition, thanks to the generated modality, the segmentation network can utilize more data information to achieve the promising results when one or more modalities are missing.

\section{Discussion and Conclusion}
\label{sec6}

In this paper, we propose a novel end-to-end feature-enhanced generation and multi-source correlation based deep neural network for brain tumor segmentation with missing MR modalities. We address the task of segmenting brain tumor in multi-parametric MR sequences, with a focus on a setting where individual modalities are missing. The proposed network architecture first generates a feature-enhanced missing modality. Then the proposed correlation constraint block is used to discover the multi-source correlation between modalities. We suppose that the correlation is linear. The multi-source correlation, on the one hand, can constrain the generator to generate the feature-enhanced missing modality; and on the other hand, can help the segmentation network to learn the correlated feature information to improve the segmentation performance. Finally, a segmentation decoder is used to achieve the brain tumor segmentation which is a multi-class segmentation.

Even if most existing synthesis approaches are based on GAN. Otherwise, it is difficult to put the generation of a 3D image by GAN and the image segmentation in a same architecture. In the literature, these two tasks are performed separately. In this paper, we include both tasks in the same framework that allows to optimize both the generation and the segmentation. In addition, the training of GAN is challenging, such as mode collapse, non-convergence, instability, and highly sensibility to hyper-parameters. Another reason that drives us not to use GAN is that, our principle goal is to utilize the multi-source correlation to segment the brain tumor when individual modalities are missing, not to generate a perfect missing modality. Therefore, we design a specific multi-encoder based network to generate the missing modalities.

To investigate the importance of the proposed components of our network, several comparison experiments are implemented with regard to the feature-enhanced generator and the correlation constraint block. The experimental results demonstrate the proposed feature-enhanced generator enable the network to generate a relevant feature-enhanced missing modality, and the correlation constraint block can aide the network to achieve the promising segmentation results. In addition, the comparison results on BraTS 2018 dataset show that out method can outperform the state-of-the-art methods despite missing modalities. 

There are several advantages of our proposed method. (i) The architecture are trained by an end-to-end fashion and fully automatic without any user interventions. (ii) The segmentation results evaluated on two metrics (Dice similarity coefficient and Hausdorff distance) are promising. (iii) The experimental results demonstrate the effectiveness of the proposed components and also our proposed method can outperform the state-of-the-art methods. (iv) The proposed method can be generalized to other kinds of multi-source images if some correlation exists between them.

However, there are some limitations of the proposed method. (i) The proposed method is only validated on multi MR modalities. (ii) There are still some rooms to improve the proposed correlation constraint block. (iii) The proposed segmentation network is evaluated on the public brain tumor dataset.

To overcome the above limitations, in our future work, we would like to expand our method to other multi-modal segmentation problem, such as CT and MRI. And it would be interesting to compare different multi-source correlation expressions (e.g. nonlinear) and to improve the performance of the method. In addition, we plan to validate our method on other public or private multi-modal segmentation datasets to investigate the robustness of our method. 

\section*{Acknowledgments}
This work was co-financed by the European Union with the European regional development fund (ERDF, 18P03390/18E01750/18P02733) and by the Haute-Normandie Regional Council via the M2SINUM project. This work was partly supported by the China Scholarship Council (CSC).

\bibliography{mybibfile}

\end{document}